\documentclass[letterpaper]{article} 
\usepackage{aaai24}  
\usepackage{times}  
\usepackage{helvet}  
\usepackage{courier}  
\usepackage[hyphens]{url}  
\usepackage{graphicx} 
\usepackage{natbib}  
\usepackage{caption} 
\usepackage{amsmath}
\usepackage{amssymb}
\usepackage{algorithm}
\usepackage{multicol}
\usepackage{multirow}
\usepackage{algorithmicx}
\usepackage{algpseudocode}

\usepackage{newfloat}
\usepackage{listings}
\DeclareCaptionStyle{ruled}{labelfont=normalfont,labelsep=colon,strut=off} 
\floatstyle{ruled}
\newfloat{listing}{tb}{lst}{}
\floatname{listing}{Listing}
\title{Attacking Transformers with Feature Diversity Adversarial Perturbation}
\author{
    Chenxing Gao\textsuperscript{1},
      Hang Zhou\textsuperscript{1},
      Junqing Yu\textsuperscript{1},
      YuTeng Ye\textsuperscript{1},
      Jiale Cai\textsuperscript{1},
      Junle Wang\textsuperscript{2},
    Wei Yang\textsuperscript{1}\thanks{indicates corresponding author.}
}
\affiliations{
    \textsuperscript{1}Huazhong University of Science and Technology, Wuhan, China\\
    \textsuperscript{2}Tencent\\

   \{cxg, henrryzh,yjqing, yuteng\_ye, jaile\_cai, weiyangcs\}@hust.edu.cn; 
    wangjunle@gmail.com;
}

\usepackage{bibentry}

\begin{document}

\maketitle

\begin{abstract}
Understanding the mechanisms behind Vision Transformer (ViT), particularly its vulnerability to adversarial perturbations, is crucial for addressing challenges in its real-world applications. Existing ViT adversarial attackers rely on labels to calculate the gradient for perturbation, and exhibit low transferability to other structures and tasks. In this paper, we present a label-free white-box attack approach for ViT-based models that exhibits strong transferability to various black-box models, including most ViT variants, CNNs, and MLPs, even for models developed for other modalities. Our inspiration comes from the feature collapse phenomenon in ViTs, where the critical attention mechanism overly depends on the low-frequency component of features, causing the features in middle-to-end layers to become increasingly similar and eventually collapse. We propose the feature diversity attacker to naturally accelerate this process and achieve remarkable performance and transferability.
\end{abstract}

\section{Introduction}

The Vision Transformer (ViT) has gained significant attention in recent years for its outstanding performance in various computer vision tasks such as object detection~\cite{beal2020toward,carion2020end}, semantic segmentation~\cite{strudel2021segmenter,zhu2021unified}, human pose estimation~\cite{xu2022vitpose,zheng20213d}, scene understanding~\cite{wu2022p2t,said2023scene} and etc. Understanding the mechanisms behind ViT, particularly its vulnerability to adversarial perturbations, is crucial for addressing challenges in its real-world applications. In general, adversarial attacks involve crafting small perturbations to input data to cause models to produce incorrect results with high confidence while being imperceptible to the human visual system.
Existing adversarial attack approaches can be categorized as ``white-box" or ``black-box" based on whether the attacker has access to the network parameters. While white-box attacks require additional knowledge about network architecture and trained weights, they are valuable because perturbations obtained from a white-box model have the potential to be transferred to attack a black-box model, i.e., possessing cross-model transferability.

Initially, adversarial attacks were studied on Convolutional Neural Network (CNN)-based models. The most commonly used methods are gradient-based approaches such as FGSM~\cite{goodfellow2014explaining}, BIM~\cite{kurakin2016adversarial}, and their variants MI~\cite{dong2018boosting}, DI~\cite{xie2019improving}, NI~\cite{lin2019nesterov} , to perturb the entire image~\cite{kim2020torchattacks} or a small patch~\cite{gao2020patch} in the same direction as the gradient of the cost function with respect to the data. Other approaches use optimization-based schemes that seek to minimize the distance between the original samples.

With the increasing prevalence of ViT-based models, their sensitivity to adversarial attacks has become a hot research area. 
Recent studies have found that ViT-based models are still vulnerable to specially crafted perturbations~\cite{hatamizadeh2022gradvit, shi2021decision, fu2022patch, lovisotto2022give}. 
%
Shi ~\cite{shi2021decision,brown2017adversarial} propose Patch-wise Adversarial Removal (PAR) black-box method to attack ViT. Wei~\cite{wei2022towards} propose to skip the gradient on the attention during the backward propagation for boosting transferability.
%
Their performances on other black-box models like CNNs and MLPs are still poor due to the structural gap. Further, most existing methods for adversarial attacks depend on classification labels to determine the cost function and compute the gradient direction. This can impede transferability because the targeted black-box model may use different category divisions, or even operate in a separate domain, from the white-box model.
%
To the best of our knowledge, the only notable exception is FG-UAP~\cite{ye2022fg}, which attacks CNN models at the layers where neural collapse occurs.
However, this method is limited to CNN-based models and focus on neural collapse layers, i.e., the last several layers.

\begin{figure*}[t]
\begin{center}
   \includegraphics[width=0.643\linewidth]{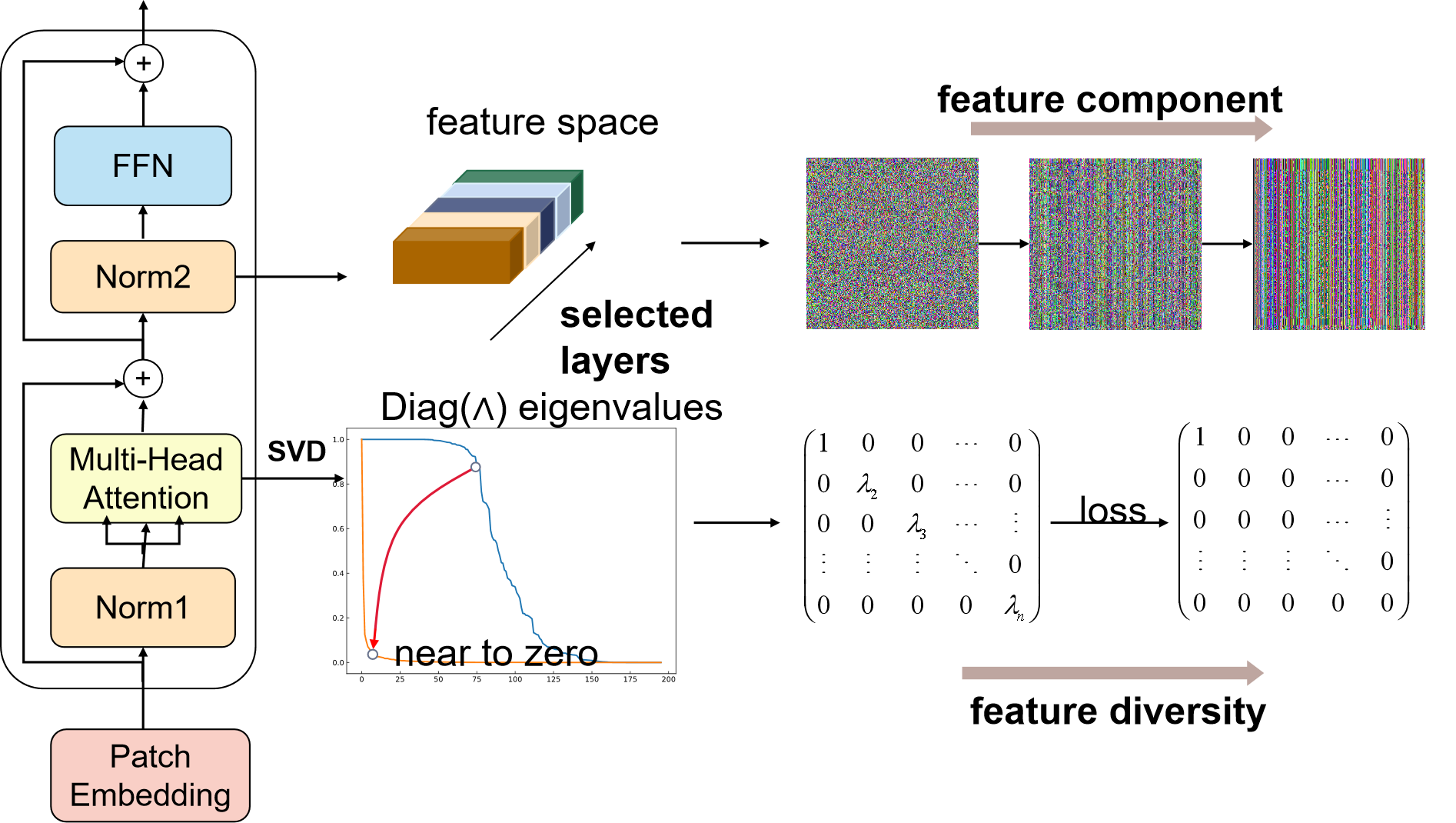}
\end{center}
   \caption{Overview of Feature Diversity Adversarial Perturbation attack (FDAP). Our attack aims to reduce the feature diversity . The reduction of the feature diversity acts as the decrease of the high-frequency components in feature space while theoretically the attention matrix gradually convergences to a rank-1 matrix.}
\label{fig:pipeline}
\end{figure*}
To address the aforementioned issues, we propose a novel label-free white-box attack approach for ViT models that exhibits strong transferability to various black-box models, including most ViT variants, CNNs, and MLPs. Our method is inspired by the analysis of the feature collapse phenomenon in ViT-based models~\cite{dong2021attention,wang2022anti}. The attention mechanism, which is a critical component for information aggregation in ViT models, is overly dependent on the low-frequency component of features, thus acting as a low-pass filter~\cite{dong2021attention}. Therefore, the features in middle-to-end layers of ViTs can become increasingly similar and eventually collapse if layers increase continuously, leading to the feature collapse phenomenon.

In this paper, we propose to attack ViT models by accelerating the feature collapse process. We select the most vulnerable layers, i.e., those with relatively high feature diversity, in transformers using the full name of Centered Kernel Alignment (CKA)~\cite{raghu2021vision}. Then, we use the high-frequency component as an indication of feature diversity and design the cost function accordingly. We name our method as Feature Diversity Adversarial Perturbation (FDAP), as shown in Figure~\ref{fig:pipeline}. The high-frequency component is a suitable indicator of feature diversity as it captures high-frequency details that differentiate features, as opposed to low-frequency details, which are commonly shared among different features. Finally, we use the positive gradient to perturb adversarial signals. 
Due to the crucial role of the attention mechanism in ViT-based models, our method exhibits broad transferability. We evaluate our method on eight ViT-based models, two CNN models, and two MLP models and demonstrate that our FDAP exhibits excellent transferability not only for ViTs but also for CNN and MLP models compared with other attack methods. Furthermore, our FDAP uses a label-free attacking scheme, allowing it to be used for attacking ViT models in other domains, resulting in cross-task transferability during experiments, as shown in Sec.~\ref{sec:cross_task}.

\section{Related work}
\noindent\textbf{Feature attacks on CNN}
In a typical attack, the loss function is generally cross-entropy loss, which suggests that the attack method is closely linked to the classification label. However, these label-related methods may not transfer well across different models. To improve transferability, some researchers have explored attacks in the feature domain. There are two main categories of feature attacks: iterative optimization approaches and generator-oriented approaches. Inkawhich et al. \cite{inkawhich2019feature} set the loss as the distance between the source image and the adversarial image activations at some layer. Yao et al. \cite{zhu2022toward} proposed an attack that drives the data distribution in feature space to deviate from the original samples. Wu et al. \cite{wu2020boosting} selected the representative feature via attention heatmap as the regularization term for the loss.

\noindent\textbf{Attacks on ViTs}
 For attack specified for ViTs, research has focused primarily on the attention mechanism, which is essential for the global modeling ability of ViT models. Hatamizadeh et al.~\cite{hatamizadeh2022gradvit} proposed the Dot Product Attention (DPA) attack, which involves reconstructing attention components to render ViT models ineffective in extracting meaningful information. Naseer et al.~\cite{naseer2021improving} suggested the Self-Ensemble (SE) method, which reorganizes the entire ViT model into an ensemble of networks to enhance attack transferability. Furthermore, the Token Refinement (TR) method~\cite{naseer2021improving} was implemented to fine-tune the class token and improve attack transferability. Additionally, because the image is in the form of patches when processed by transformer blocks, some attacks perturb only a few patches that significantly affect classification, such as the Patch-Fool attack~\cite{fu2022patch}. Architecture-oriented Transferable Attacking
(ATA) framework~\cite{wang2022generating} activated uncertain attention and perturbed the sensitive embedding. Token Gradient Regularization (TGR) method~\cite{zhang2023transferable} utilized the regularized gradient to generate transferable adversarials.

\noindent\textbf{Feature collapse}
Feature collapse is observed in DeepViT~\cite{zhou2021deepvit}, which means that feature map of ViT tend to be similar when transformer blocks become deeper. Dong  \cite{dong2021attention} were the first to give a theoretical explanation of feature collapse in ViT. Diversity constraints, in general, seek to learn discriminative patterns for feature coverage, such as cosine similarity-based, distance-based, orthogonality-based, and so on~\cite{gong2021vision,chen2022principle}. To some extent, feature diversity is positively correlated with ViT classification accuracy. 

\section{Methodology}
 
For a given classifier $\mathcal{F}$ to be attacked: $\mathcal{F}:X \xrightarrow{}Y$, where $X$ represents the input space and $Y$ represents the label space accordingly. Randomly choose an image
$x \in X$ and its corresponding ground truth $y \in Y$. The goal is to find an optimal perturbation $\delta$ to $x$ that leads to misclassification, while preserving visual resemblance. In general, a gradient-based attack problem is as:
\begin{equation}
\delta \leftarrow \arg \underset{\delta}{\max }\; J\big ( \mathcal{F}(x+\delta), y  \big ), \,\,\,\,
\text{s.t.}\;\left\|\delta\right\| \leq \epsilon
\label{equ:definition}
\end{equation}
\noindent where $\epsilon$ is the upbound of the perturbation $\delta$. Typically, the loss $J$ is set to the cross-entropy loss.

\subsection {Feature Diversity Perturbation}
Feature diversity~\cite{tang2021augmented}, which indicates the degree of a set of features within a layer diversify, plays a crucial role in the performance of ViTs. The diversity of  a feature map $z$ can be measured as the distance between $z$ and its nearest rank-1 matrix $\mathbf{1} z_{0}^{T}$. 

\begin{equation}
r(z) = \left\|z-\mathbf{1} z_{0}^{T}\right\|_F 
\label{eqdiver}
\end{equation}
\noindent where $\mathbf{1}$ is the all-ones vector, and $z_{0} = \arg \underset{z_{*}}{\min } \left\|z- \mathbf{1} z_{*}^{T}\right\|$, thus the rank of $\mathbf{1} z_{0}^{T}$ is 1.

A larger diversity score indicates a more diverse set of features within a layer. According to the proof in ~\cite{wang2022anti}, the nearest rank-1 matrix $\mathbf{1} z_{0}^{T}$ is exactly the Direct Component (DC) of feature $z$. Hence, we argue that the measure of feature diversity is actually the norm of residual components (after removing the DC) of feature $z \in \mathbb{R}^{n \times d}$. In the following, we show the rationality of such a definition. 
 \begin{equation}
     \left\|z-\mathbf{1} z_{0}^{T}\right\|_F^{2}=nz_{0}^{T}z_{0}-2\text{Tr}(z^{T}\mathbf{1} z_{0}^{T})+\text{Tr}(z^{T}z)
\label{eq2}
 \end{equation}
\noindent where $n$ is the number of features, its derivative can be denoted as:
\begin{equation}
  \nabla_{z_0} \left\|z-\mathbf{1} z_{0}^{T}\right\|_F^{2}=2nz_{0}-2z^{T}\mathbf{1}
  \label{eq:z_deriv}
\end{equation}
Therefore, the value of Eqn.~\ref{eq:z_deriv} must be zero for the rank-1 matrix $z_{0}$ satisfying Eqn.~\ref{eqdiver}. Then we can conclude that $z_{0}$ is the direct component of feature $z$:
\begin{equation}
    z_{0}^{T} = \frac{1}{n}\mathbf{1}^{T}z
\label{eqdc}
\end{equation}
\noindent Hence, the feature diversity actually measures the alternating components (high-frequency components) of the feature. While low-frequency components emphasize global information, high-frequency components offer more detailed and local information. 
The corresponding diversity for the feature map $f_k(x)$ for input $x$ at layer $k$ is $r(f_k(x))$.  We define our feature diversity loss for adversarial attacking as:
\begin{equation}
    J_{\text{FD}}(x)=-\sum\limits_{k\in \mathcal{S}}[\text{log}(r(f_k(x+\delta))]^{\beta}
\label{eqJ}
\end{equation}
 
\noindent where $\mathcal{S}$ is the set of selected layers to attack, $\beta$ is the accelerating parameter for simulating the natural feature collapse process in ViT.
The adversarial problem is correspondingly transformed into:
\begin{equation}
 \delta \leftarrow \arg \underset{\delta}{\max }\; J_{\text{FD}}((x+\delta)) , \
  \text{s.t.} \; ||\delta|| \leq \epsilon 
\label{eqestablish}
\end{equation}
\noindent Notice our formulation doesn't rely on the groundtruth label $y$.

\subsection{Attacking ViT with Feature Diversity}

In the above section, we define our feature diversity attach formation. However, directly applying it in ViT triggers practical questions, i.e., the transformer layers contain various modules, and which feature we should attack? We dive into the frequency component of each module within the transformer layer and analyze the frequency responses. 
A transformer layer is consisting of a multi-head self-attention mechanism (MHSA), a skip connection (SC) and a fully connected feed-forward network (FFN). In the following, we analyze the frequency response of each component.

The MHSA relies on query $Q$, key $K$ for re-weighting value $V$. To analyze the frequency change w.r.t. the input after MHSA module, naively we can conduct Discrete Fourier Transform (DFT) by solving a least squares problem $\underset{\alpha_{j}}{\min}\sum\limits_{i}|\sum\limits_{j} \alpha_{j}F_{ij}-ff_{i}|^2$ to obtain Fourier coefficients $\alpha_{j}$, where $F_{ij}$ are the real and imaginary parts and $ff_j$ denotes the Fourier basis.
%
The above optimization problem is ill-conditioned for high-dimensional features~\cite{barannyk2015spectrally,huybrechs2010fourier}.
Instead, we turn to Singular Value Decomposition (SVD). 
The target is to explore the change of frequency components instead of the actual coefficients. What we need to figure out is how the frequency components are affected in MHSA operation. And SVD can offer us the influence of attention on the frequency components by analyzing the eigenvalues of the corresponding frequency components.

For a transformer block for ViT, we show how MHSA decompose feature using SVD as followed:
\begin{equation}
\begin{split}
    A & = \text{softmax}(\frac{QK}{\sqrt{d}}) \\
    &= [u_1,u_2,\cdots,u_n] \Lambda [v_1^{T},v_2^{T},\cdots,v_n^{T}]
\label{eqsvd}
\end{split}
\end{equation}
Then we record the value $V$ as:
\begin{equation}
 V= [x_1,x_2,\cdots\,x_n]
 \label{eqQ}
 \end{equation}
where $A$ denotes the attention matrix after softmax operation. 
\begin{equation}
\begin{split}
    Ax_i & = (u_1,u_2,\cdots,u_n)\Lambda(v_1^{T},v_2^{T},\cdots,v_n^{T})x_i \\
    &= (\lambda_{1}u_{1}v_1^{T}+\lambda_{2}u_{2}v_2^{T}+\cdots+\lambda_{n}u_{n}v_n^{T})x_i
\label{eqsvd_2}
\end{split}
\end{equation}
\noindent where $\lambda_1$, $...$, $\lambda_n$ are the eigenvalues of the matrix $A$ for Singular Value Decomposition.
This shows the eigenvalues and its corresponding eigenvectors re-weight the feature. With the softmax activation, the eigenvalues are limited to the range of (0,1] (See the detailed proof in Appendix.)
The eigenvalues illustrate the remaining degree of each feature through an attention mechanism. eg if a feature component responds to an eigenvalue of 0.9, then the component will be reduced to $90\%$ of its original value after the attention mechanism. The feature component with the maximum eigenvalue of 1 corresponding to the direct component, and all other components are weakened in comparison.

In the SC module, it is a supplement of the original feature by adding MHSA output to the original feature, which is expressed as:
\begin{equation}
    Z = \mathrm{MHSA}(X)+X
\label{eqsk}
\end{equation}
where $X$ denotes the feature after the first norm layer (norm1). The subsequential Feed-Forward Network(FFN) module is likely to migrate the feature collapse with an upper bound $\sigma$ \cite{dong2021attention,wang2022anti} to both amplify the high-frequency component and the low-frequency component.

In FFN, only the resnet component is processed, attack the FFN resnet is not that effective compared with attacking the bone. We denote FFN operation of $X$ as $P(X)$. Then we consider $\nabla_{P}\mathcal{F}$, where $\mathcal{F}$ denotes the final classification loss. With chain rule, we have:
\begin{equation}
\nabla_{P}\mathcal{F}= \nabla_{X}\mathcal{F}\nabla_{P}X=\nabla_{X}\mathcal{F}/\nabla_{X}{P} \\
\label{equder_p}
\end{equation}
\noindent We notice that the difference between the $\nabla_{P}\mathcal{F}$ and 
 $\nabla_{X}\mathcal{F}$ is just a parameter $1/\nabla_{X}P$. Since FFN amplify its element with the same degree,$\nabla_{X}P$ is larger than 1 \cite{wang2022anti}, therefore $\nabla_{P}\mathcal{F}$ is less than  $\nabla_X\mathcal{F}$. The derivation indicates that changing $X$ is more effective than FFN module $P(X)$. 
\begin{figure}[t]
\begin{center}
   \includegraphics[width=0.8\linewidth]{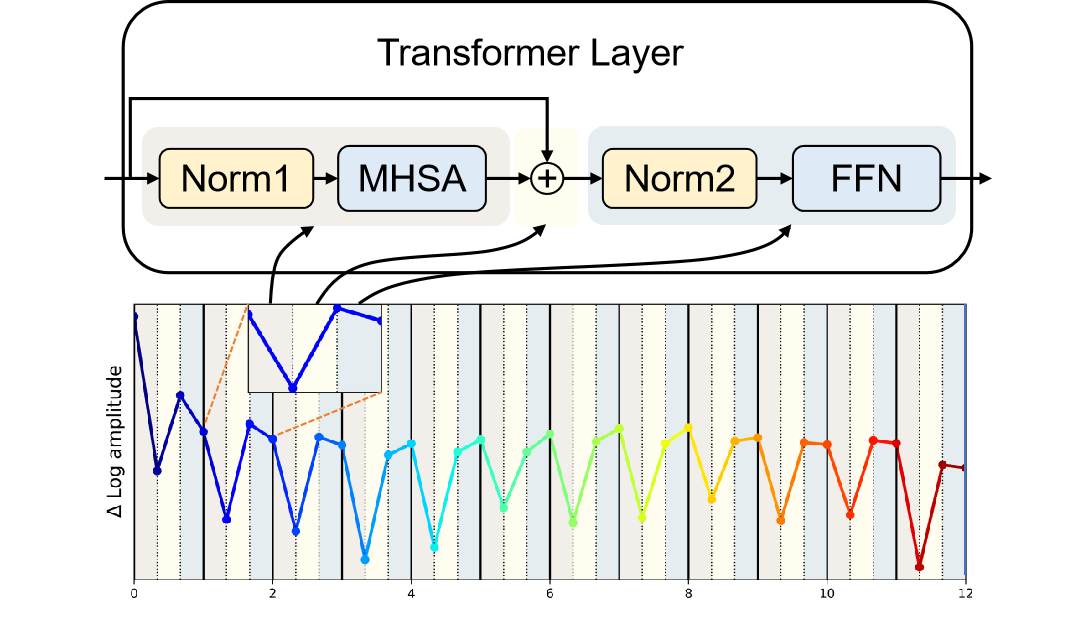}
\end{center}
   \caption{Frequency change due to different modules of a ViT model. The x-axis represents different modules of transformers blocks while the y-axis represents the the delta of the frequency. The light-purple, light-yellow, and blue columns respond to the MHSA, SC and FFN module respectively. The decrease of the value of y represents the low-pass filter. }
\label{fig:freq}
\end{figure}
Recall feature diversity aims to reduce the high frequencies within features. Considering the above analysis, features after MHSA contains low-frequency components. However, attacking FFN module is not ideal since the final result is not sentitive to the change of the FFN compared with the main branch feature. Besides, features after the Skip connection have not been normalized, thus, the distribution of features at this point has a gap with the feature suitable for the model. In summary, attacking the second layer norm (norm2) is the best choice. As a result, the corresponding diversity loss is:
\begin{equation}
J_{FD}=-\sum\limits_{k=i}^j[\text{log}(r(\text{norm2}_k(x+\delta))]^{\beta}
\end{equation}
\begin{algorithm}[tb] 
       \caption{Feature Diversity Adversarial Perturbation on ViTs}
       \label{alg}
       \textbf{Input}: The feature diversity loss $ J_{FD} $, the clean sample 
		$ x $, the surrogate model $ f $, the maximum perturbation $\epsilon$, the number of iterations $N$.\\
       \textbf{Output}: The adversarial image $x^{\text{adv}}$.\\
       \textbf{Initialize}: $\delta_{0}=0$, $g_{0}=0$, $\alpha=3/255$
       \begin{algorithmic}[1] 
        \For{t = 0 to N-1}
		\State $g_{t+1} = \nabla_{\delta}J_{\text{FD}}(x+\delta_t)$
        \State $\delta_{t+1} = \alpha \cdot \operatorname{sign}\left(g_{t+1}\right) $ 
		\State $x^{\text{adv}}_{t+1} = {\text{Clip}}_{x,\epsilon}\left\{x^{\text{adv}}_t+\delta_{t+1} \right\} $
		\EndFor
        \State {\bf return}   $  x^{\text{adv}}_{N}$
		
	\end{algorithmic} 
\end{algorithm}
where $\text{norm}2_k(x+\delta)$ represents the feature of perturbed image $x+\delta$ after the second layer norm operation at layer k.
Figure \ref{fig:freq} provides an illustration of the frequency changes resulting from various mechanisms applied to 12 transformer blocks for a randomly selected image. The figure reveals that the MHSA mechanism functions as a low-pass filter, while its accompanying Skip Connection acts as a corresponding high-pass filter. These observations suggest that, in general, the high-frequency components of features are weakened in transformer blocks.

 %
The attack on feature diversity effectively reduces high-frequency feature components for ViT, details are shown in Algorithm \ref{alg}.

\section{Layers to Attack}
The layers we apply our feature diversity attacking have a great impact on the final effectiveness. Naively attacking all layers diminishes the attacking effect as the not all the layers have significant impact on the classification. It is essential to explore the similarity of layers so that we can figure out how features transfer cross layers. Rather than attacking all layers, we select the layers according to Centered Kernel Alignment ($\mathrm{CKA}$). $\mathrm{CKA}$ is a similarity criterion for quantitative comparisons of representations. For features of two different layers $ f_i(x) \in \mathbb{R}^{n \times p_{1}}$, $f_j(x) \in \mathbb{R}^{n \times p_{2}}$, we first calculate the measure of statistical independence named as Hilbert-Schmidt independence criterion($\mathrm{HSIC}$), formulated as:
\begin{equation}
    \mathrm{HSIC}(X, Y) = \mathrm{vec}(X') \cdot \mathrm{vec}(Y')/(n-1)^2
    \label{eqhsic}
    \end{equation}
\noindent where $X=f_i(x)f_i(x)^T$, $Y=f_j(x)f_j(x)^T$ are the Gram matrix for the feature of two layers, $X' = HXH$ and $Y' = HYH$ are corresponding centered Gram matrices, the centering matrix $H=I_n-\frac{1}{n}11^{T}$. The larger the value of CKA score is, the more similarity the feature of different layers shares.
\begin{align}
    \mathrm{CKA}(X, Y) = \frac{\mathrm{HSIC}(X, Y)}{\sqrt{\mathrm{HSIC}(X, X) \mathrm{HSIC}(Y, Y)}}
\label{eqcka}
\end{align}
\begin{figure}
\begin{center}
   \includegraphics[width=0.7\linewidth]{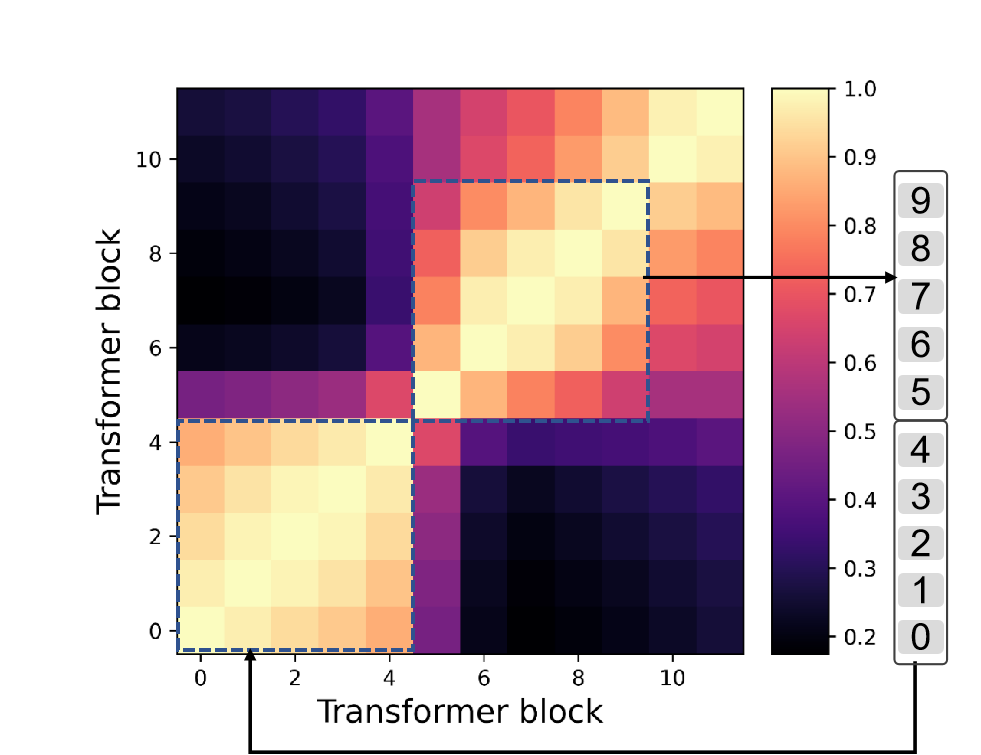}
\end{center}
   \caption{Layer Selection with CKA analysis. More yellowish indicates more similarity between two layer features. We can see that there are three block structures. We select the layers in the second block because in these layers features show relatively high feature diversity. }
\label{fig:cka}
\end{figure}
In Figure~\ref{fig:cka}, we exhibit the CKA of 50 clean samples in the DeiT-B-Distilled model, in which there are three obvious blocks. In the initial layers (0-4 in Figure~\ref{fig:cka}), the features don't contain enough information for models to classify. In the middle layers, the components of the features are shifted at a large scale, transformer-blocks begin to filter high-frequency components at the stage. Towards the end of the layers, the features are designed to fit the final MLP head. Thus, the features at these layers are too model-specific and don't share good transferability cross models. Besides, for the layers after the first block, the similarity maintain a relatively high level (above 0.5) compared with the initial layers(nearly to 0.3), indicating that the features after the first block are transferred cross layers at a large scale. 
Based on this analysis, we choose to target the middle layers indicated by the second block.

\section{Rationality of Transferability}
For strong transferability, a key factor is the feature's high contribution to the task, independent of the model structure. Such features can be disrupted to have the most significant impact on the final result. In this section, we analyze the transferability of our FDAP attack using this perspective.
We first focus on ViTs and identify high-frequency components as the critical feature. Although ViT-based models have different structures, they all use similar attention and skip connection mechanisms in their transformer blocks. DeiT adds a distillation token, resulting in changes to Eqn.~\ref{eqsvd}, Eqn.~\ref{eqQ} and Eqn.~\ref{eqsvd_2}. ConViT is divided into local and nonlocal stages, where the attention is combined with CNN and ViT, and $\sigma(\lambda)$ determines the percentage of engagement by CNN. We learn that $\sigma(\lambda)$ is greater than 0 only in early layers. In CaiT, the learnable diagonal matrix across feature channels reduces feature collapse in the MHSA stage, but it still occurs in the last few layers. Class-attention employs a softmax operation, resulting in decreased feature diversity.
For CNNs, our attack weakens informative high-frequency components, essential for their operation. For MLPs, the Mixer layer replaces the attention with MLP(FFN) modules, resulting in more high-frequency components being passed through than in ViTs.

\section{Experiments}
We conduct experiments to test our method on ViT-based models, CNN models, and MLP models. Furthermore, we  assess the cross-task transferability of our attack method. We follow the white-box attacking protocol, i.e., generate the adversarial sample on white-box models and test on black-box models.

\subsection{Experimental Setup}
\noindent\textbf{Dataset:}~Similar to the settings in Dong\cite{dong2018boosting}, we randomly select data 1000 images from the validation set ImageNet 2012 \cite{russakovsky2015imagenet}. Each image is selected from corresponding class and can  be almost correctly classified by all baselines for evaluation.

\begin{table*}[htbp]
\renewcommand{\arraystretch}{2}
\begin{center}
{
              \fontsize{9pt}{1pt}\selectfont
                \setlength{\tabcolsep}{2pt}
            {
            \begin{tabular}{c|c|cccccccccccc}
            \hline
            model                       & method &ViT-L  &DeiT-Dis & CaiT-S  &ConViT-B  & DeiT-B    & CaiT-XXS & ConViT-S      & TNT-S &Res50 &Res101 &Mixer-B &Resmlp-24\\ \hline
            \multirow{10}{*}{ViT-L/16}         & MIM      & 99.7  & 34.1     & 37.7   & 42.4    & 48.4      & 47.0     & 46.1          & 28.4  & 35.0     & 24.2      & 35.1       & 53.5       \\
                                        & DIM      & 99.9  & 34.7     & 38.1   & 44.1   & 48.1      & 47.7     & 46.9          & 31.7  & 32.3     & 24.9      & 36.1       & 54.1      \\
                                        & NIM      & 100   & 34.5     & 35.4   & 29.3       & 47.5      & 46.5     & 47.2          & 31.7  & 34.7     & 26.8      & 36.2       & 53.3       \\
                                        & Patch-Fool & 99.9  & 15.8     & 16.6   & 25.2      & 41.3      & 26.2     & 38.2         & 16.9 
                                        & 38.9     & 30.6     & 26.6       & 35.6 \\
                                        & SE      & 99.6  & 14.6     & 13.6   & 15.9      & 13.3      & 20.8     & 16.4          & 13.0  & 20.5     & 14.5      & 14.9       & 21.5         \\
                                        &SAGA     & 99.9  & 12.5     & 12     & 14.5    &  12.4      & 21.6    &15.6
                                        &  15.3       &   24.8     &18.2     & 12
                                        & 21.1     \\
                                        & ATA   & 99.9  & 13.9     & 11.2   & 20.7 & 19.1      & 23.2     & 23.9         & 13.0           & 20.5     & 14.5      & 14.9       & 21.5 \\
                                         & PNA     & \textbf{100}   & 41.9     & 45     & 40.1     & 53.8     & 50.1     & 52.8          & 40.1 &  36.7     & 28.6      & 52.4       & 53.3  \\    
                                        & Ours    & 99.9  & 54.7     & 55.6   & 50.3    & 56.5      & 69.0     & 59.0          & 40.2  & 37.3     & 30.9      & 64.6       & 67.2      \\
                                       & MIM+Ours  &99.7 & \textbf{59.1}&\textbf{60.8}  &\textbf{55.4} & \textbf{59.5}      & \textbf{73.9}     & \textbf{62.7}          & \textbf{43.1}  & \textbf{40.0}    & \textbf{33.1}      & \textbf{68.3}       & \textbf{76.2}       \\ \hline
    \multirow{10}{*}{DeiT-Dis}       & MIM    & 17.6  & 99.2     & 46.3   & 53.4     & 65.0      & 52.0     & 57.5          & 41.5  & 37.3     & 25.1      & 37.0       & 54.9       \\
                                        & DIM       & 19.1  & 99.6     & 46.9   & 54.8   & 66.4      & 53.6     & 60.3          & 41.5  & 38.6     & 26.0      & 37.7       & 55.0      \\
                                        & NIM      & 17.2  & 99.3     & 27     & 42.4    & 45.9      & 36.5     & 39.0          & 28.5  & 35.0     & 21.1      & 28.9       & 42.9       \\
                                        & Patch-Fool   &  6.1  & 99.8     & 10.2   & 14.2 & 18.9      & 16.4     & 13.9          & 11.3  & 16.8     & 11.2      & 11.4       & 18.2   \\
                                        & SE      & 11.7  & 98.9     & 43.6   & 29.7     & 45.2      & 42.1     & 43.6          & 27.6  & 27.5     & 17.2      & 30.1       & 48.4   \\
                                        &SAGA     & 6.5   & 98.6     & 13.6   & 15.5    & 13.1          &21.3
                                        &   15.8      &  15.7         &24.3
                                        &    18.3        &    12      &21.6   \\
                                         & ATA     & 14.5  & \textbf{100}     & 16.8   & 31.2     & 23.5      & 23.2     & 32.5    & 22  & 20.5     & 14.5      & 14.9       & 21.5        \\
                                         & PNA     & 13.2  & 99.9     & 16.8   & 32.2   & 67.5   & 59.1     & 67.7         & 38     & 43.7    & 33.7     & 57.4       & 73.6 \\
                                        & Ours     & 20.2  & 99.4     & 62.4  & 63.7      & 68.0      & 57.5     & 66.6          & 37.9  & 45.4     & 39.4      & 58.2       & 74.3      \\
                                        & MIM+Ours  &\textbf{26.9}& 99.2&\textbf{66.7} & \textbf{68.2} & \textbf{77.2}      & \textbf{73.4}     & \textbf{77.7}        & \textbf{46.2}  & \textbf{60.9}     & \textbf{54.1}      & \textbf{70.5}       & \textbf{86.8}        \\ \hline
           \multirow{9}{*}{CaiT-S}           & MIM     & 14.8  & 23.4     & 97.8   & 41.3        & 48.9      & 42.2     & 42.1          & 32.5  & 34.8     & 20.1      & 28.2       & 43.3       \\
                                        & DIM       & 15.3  & 25.2     & 98.7   & 41.3   & 50.3      & 43.7     & 43.7          & 34.9  & 35.9     & 22.2      & 30.7       & 43.4    \\
                                        & NIM      & 13.7  & 24.1     & 98.3   & 40.7    & 39.9      & 33.0     & 34.0          & 26.6  & 34.1     & 19.5      & 25.5       & 39.5    \\
                                        & SE      & 13.1  & 31.8     & 97.1   & 33.7   & 37.2      & 41.5     & 37.5          & 26.5  & 26.3     & 17.0      & 25.5       & 43.1    \\
                                        &Patch-Fool &7.1    &9.6   &97.2  &12 &13.4 &19.5 
                                        &19 &12.2  &25.1   &16.9 &16.9 &22.5 \\
                                        &SAGA     & 9.6   & 13.7     & 98.7   & 15.9      & 12.7      & 22 & 16.5       & 16.3      & 24.5
                                        &19.1        & 11.2  &21\\  
                                        &ATA     &15.1     &31.9    &98.5     &13 &43.2
                                        &38.2   &45.5  &13.6 &29.2 &15.8 &42.3 &43.6 \\
                                         & PNA       & 15.8  & 49.4     & \textbf{99.8}   & 45.4  & 55      & 60.3     & 55.2         & 35.1    & 39.4     & 30.5      & 46.9      & 70.8 \\       
                                        & Ours   & \textbf{16.7}  & 50.3     & 99.6   & 45.5     & 54.7      & 62.1     & 56.0          & 35.3  & 40.1     & 34.8      & 48.7       & 71.8       \\
                                        & MIM+Ours  &16.5 &\textbf{63.1}&99.5 &\textbf{50.2}  & \textbf{61.0}      & \textbf{69.1}     & \textbf{63.9}          & \textbf{38.1}  & \textbf{46.4}     & \textbf{40.6}      & \textbf{58.4}      & \textbf{79.9}       \\ \hline
            \multirow{9}{*}{ConViT-B}         & MIM     & 15.7  & 13.6     & 20.8   & 95.4       & 52.2      & 35.6     & 47.9          & 29.3  & 32.9     & 18.2      & 29.1       & 42.5      \\
                                        & DIM    & 16.5  & 25       & 31.1   & 96.6    & 53.2      & 38.0     & 50.3          & 32.0  & 34.5     & 19.7      & 30.8       & 44.0     \\
                                        & NIM      & 15.3  & 20.1     & 28.9   & 96.3   & 41.2      & 30.6     & 36.1          & 24.8  & 32.4     & 18.1      & 26.4       & 36.3     \\
                                        &Patch-Fool &7.5     & 10.2    &9.3  &96.1 &15.6
                                        &20.1     &20.3      &16.8     &25.1  &23.6  & 24.1
                                        &26.7\\
                                        & SE     & 18.9  & 42       & 39.8   & 98.7    & 56.4      & 45.4     & 61.2          & 36.7  & 27.1     & 22.7      & 33.7       & 46.1     \\
                                        & SAGA    & 7.5   & 13.4     & 13.6   & 95.8   & 12.2     & 21  &  16.7      &  15.2    & 24.4
                                        & 17.8       & 11.3    &21.3 \\
                                        &ATA &18.3 &18.6   &16.3 &98.9 &23.6 &27 &27.9 &9.2 &26.5 &14.2 &32.3 &49.5 \\
                                        &PNA   & 19.3  & 58.1     & \textbf{71.2}   & \textbf{99.8}     & 67.7      & 61.7     & 78.6          & 41.7     & 39.3     & 23      & 61.8      & 76.3  \\
                                        & Ours    & 21.7  & 62.3     & 67.5   & 99.1   & 67.6      & 65.4     & 77.7          & \textbf{45.5}  & 47.2     & 42.5      & 64.5       & 79.7       \\
                                        & MIM+Ours  &\textbf{24.1} &\textbf{74.1} &69.5 &99.4  & \textbf{69.1}      & \textbf{67.8}     & \textbf{78.7}          & 45.1  & \textbf{49.1}     & \textbf{44.7}      & \textbf{65.5}       & \textbf{80.3}       \\ \hline
            \end{tabular}
}
}
\caption{The fooling rate of 1000 adversarial samples generated by ViT-based white-box models for different black-box models, including ViT-based models, CNN and MLP models.}
\label{tabel2}
\end{center}
\end{table*}

\noindent\textbf{Target models:}~For the comprehensiveness of evaluation, we conducted two distinct experiments. In the first experiment, we evaluated the performance of both white-box and black-box models based on the ViT architecture. The second experiment, on the other hand, involved white-box ViT models, while the black-box models consisted of CNN and MLP architectures. The ViT models include ViT-L/16, DeiT-B, DeiT-B-Distilled (DeiT-B-Dis), CaiT-S, CaiT-XXS, ConViT-B, ConViT-S and TNT-S. The CNN and MLP models include ResNet50, ResNet101, Mixer-B/16 and ResMLP-24.

\noindent\textbf{Baseline attacks:}~We evaluate our proposed method against several approaches, including SE, SAGA\cite{mahmood2021robustness}, Patch-Fool\cite{fu2022patch}, PNA\cite{wei2022towards}, ATA\cite{wang2022generating}, MI-FGSM (MIM), DI-FGSM (DIM), and NI-FGSM (NIM). While the reporter proposing SE also presents the TR method, which enhances the transferability of attacks, we do not include it in our comparison due to its requirement for fine-tuning the ImageNet dataset. 

\noindent\textbf{Attack settings:}~The value of our loss function is larger than the value of a common CE loss, hence our method requires more steps and larger step-size to converge. We tested the parameters and compare with MI, DI, and other methods in default settings. Using our parameters, the performance is improved from 0 to 0.5$\%$.
 we conduct attacks using a maximum perturbation value of $\epsilon =16$, the total number of attack iterations is $N=30$, and the step size $\alpha= 3/255$.  

For the common variants of ViT, we choose the feature layers at 5-9.  For the CaiT, we attack the feature layers at 17-20. This is because the second similarity block structure of CaiT appears after layer 17. Specifically, for ViT-L, we select the feature at 5-12. This is because the feature of ViT-L in the second similarity block is too model-specific. For the other methods, we follow the original settings of each baseline method.

\noindent\textbf{Metric:}~We use the fooling rate, the percentage of images whose labels are changed, on the whole test set to evaluate the accuracy.
\subsection{White-box Attack on ViTs}
In this section, we study the fooling rate on ViT-based models in white-box attack settings. All the results are listed in Table~\ref{tabel2}. Since our method doesn't require labels, it is slightly less superior to the other methods when the surrogate models and the victim models are the same ones. 
\subsection{Transferability to ViTs}
In this study, we focus on evaluating the feature diversity loss in attacking various ViT-based models. Specifically, we use one model as the surrogate model and the others as victim models for evaluation purposes. The results are presented in Table~\ref{tabel2}, where the surrogate model is represented along the vertical axis, and the victim models are represented along the horizontal axis. We compare the performance of our method with other baseline methods using various attacking methods.

From Table~\ref{tabel2}, we observe: (1) Our method outperforms other baseline methods in most cases.  (2) We also observe that different ViT-based white-box models show varying transferability with ViT-based models. (3) We further demonstrate that our method can be combined with other methods to achieve better performance. 

\subsection{Transferability on CNNs and MLPs}

To demonstrate the effectiveness of our proposed method in improving cross-model transferability on CNN and MLP models, we conducted an experiment using ViT-based models as white-box models and CNN and MLP models as black-box models. The results of our experiment are presented in Table ~\ref{tabel2}.

From Table \ref{tabel2}, we can conclude that:(1) Our method demonstrates superior transferability for both CNN and MLP models compared to existing methods, for the majority of cases. In particular, we observed a much higher fooling rate on MLP models compared to CNN models, which we attribute to the similarity in architecture between MLPs and ViTs (FFN layers). (2) We observe a significant enhancement in transferability for MLP models with the use of our method. For instance, the fooling rate on Mixer-B almost doubled compared to other methods when ConViT-B is chosen as the white-box model. These results suggest that our method can be used to attack common types of models, such as ViTs, CNNs, and MLPs.


\subsection{Cross-Task Transferabililty}
\label{sec:cross_task}

For label-based attackers, the target black-box model should have the same tasks and category divisions as the white-box model. While in practice, the targeted black-box model varies, which may use different category divisions, or even operate in a separate domain from the white-box model. As such, the use of labels severely limits the application scope of the method. In contrast, label-free attackers have no such issue.

In this section, we aim to demonstrate the effectiveness of our FDAP method on tasks beyond image classification, including object detection, semantic segmentation, pose estimation and depth estimation. To do so, we utilize ViT-based models, including DeiT-T, DeiT-S, and DeiT-B, as white-box models, and select common tasks and their most representative ViT-based models. Some of them achieve SOTA over a long time. To be concrete, we select the Detection Transformer (DETR) as a black-box model for object detection, SegFormer as a black-box model for semantic segmentation, ViTPose for pose estimation, and Swin-v2-MIM as a black-box model for depth estimation. 
\begin{table}[t]
\begin{center}
\resizebox{1.0\columnwidth}{!}{
 \fontsize{9pt}{1.5pt}
             \setlength{\tabcolsep}{1pt}  
\begin{tabular}{c|c|cc|cc|cc}
\hline
Target $\rightarrow$              & DETR                             & \multicolumn{2}{c|}{SegFormer}                             & \multicolumn{2}{c|}{ViTPose-B}                       & \multicolumn{2}{c}{Swin-v2-MIM}                      \\ \hline
\multirow{2}{*}{Task$\rightarrow$} & \multirow{2}{*}{\begin{tabular}[c]{@{}c@{}}Object \\ Detection\end{tabular}} & \multicolumn{2}{c|}{\multirow{2}{*}{\begin{tabular}[c]{@{}c@{}}Semantic \\ Segmentation\end{tabular}}} & \multicolumn{2}{c|}{\multirow{2}{*}{\begin{tabular}[c]{@{}c@{}}Pose \\ Estimation\end{tabular}}} & \multicolumn{2}{c}{\multirow{2}{*}{\begin{tabular}[c]{@{}c@{}}Depth \\ Estimation\end{tabular}}} \\
                                     &                                  & \multicolumn{2}{c|}{}                                      & \multicolumn{2}{c|}{}                                & \multicolumn{2}{c}{}                                 \\ \hline
Source                               & mAP                              & mIoU   & mAcc  & AP  & AR & Abs Rel$\downarrow$   & Sq Rel$\downarrow$                                            \\
clean                                & 44.03                            & 78.56  & 85.43 & 75.53                                           & 80.96                                          & 0.089                                            & 4.38                                          \\
DeiT-T                               & 15.20                             & 31.17  & 38.03 & 44.32                                           & 50.32                                          & 0.122                                            & 7.27                                          \\
DeiT-S                               & 14.80                             & 30.22  & 36.76 & 44.11                                           & 50.14                                          & 0.120 & 7.24                                          \\
DeiT-B                               & 15.50                             & 29.62  & 36.15 & 43.65                                           & 49.56                                          & 0.120 & 7.20                                           \\ \hline
\end{tabular}
}
\end{center}
\caption{Cross-Task transferability for different tasks, including
object detection, segmantic segmentation, pose estimation and
depth estimation.}
\label{table3}
\end{table}

As ViT-based models require input images of size 224$\times$224, we follow the settings in Naseer et al~\cite{naseer2021improving} and resize the clean images to new shape whose height and width are the times of 224. 

As for classification to object detection, we select the COCO 2017 dataset to demonstrate the cross-task transferability of our method. Our evaluation criterion is the degradation of mean Average Precision (mAP) .For classification to pose estimation, COCO 2017 dataset is also selected. We utilize the degradation of Average Precision(AP) and Average Recall(AR) as the evaluation matrix. For classification to semantic segmentation, Cityscapes is set as dataset, and the degradation of mIoU and mAcc are set as the evaluation. For classfication to depth estimation, we choose the val set of NYUV2 as the dataset and evaluate the attack performance with Abs Rcl and Sq Rel. All the results are listed in Table~\ref{table3}.
We notice that the performance of theses black-box models degrades at a large scale. eg, mIoU of SegFormer even drop to less than half of its original performance.

\section{Ablation Study}

\begin{figure}
\begin{center}
   \includegraphics[width=1.0\linewidth]{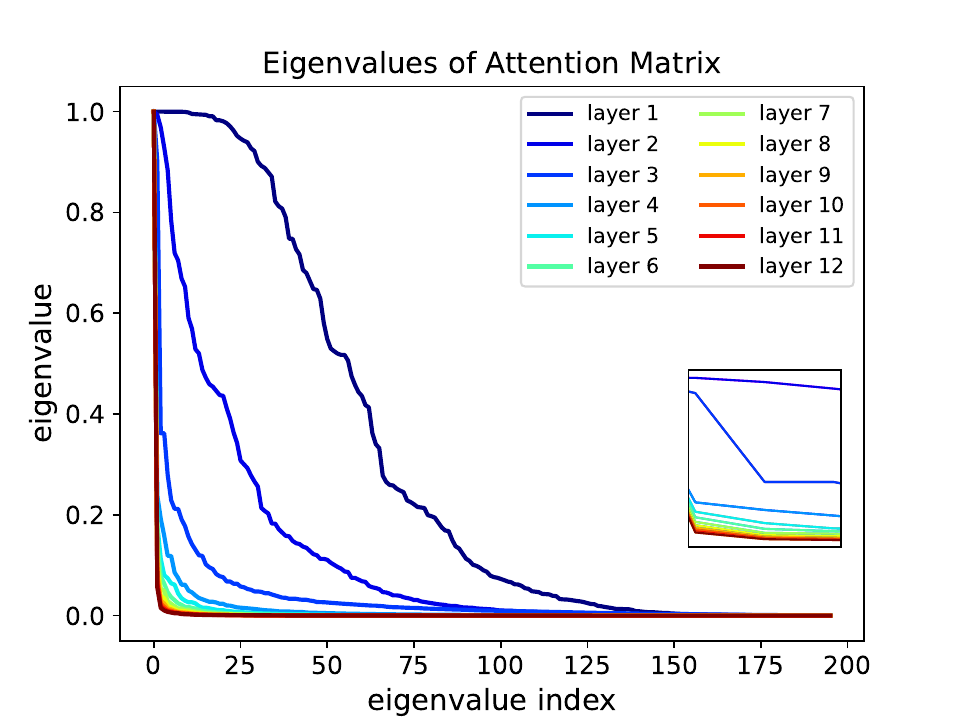}
\end{center}
   \caption{Eigenvalues for ViT attention matix of all layers.}
\label{fig:eig}
\end{figure}

 We investigate the relationship between feature layers and the corresponding eigenvalues for attention mechanisms in Figure~\ref{fig:eig}. We observe that the eigenvalues gradually converge to a distribution in which almost all the values are nearly zero, except for one that is equal to one. This phenomenon becomes more evident from layer 4, which is consistent with our layer selection at layer 5-9.

\begin{figure}[!t]
\begin{center}
   \includegraphics[width=1.0\linewidth]{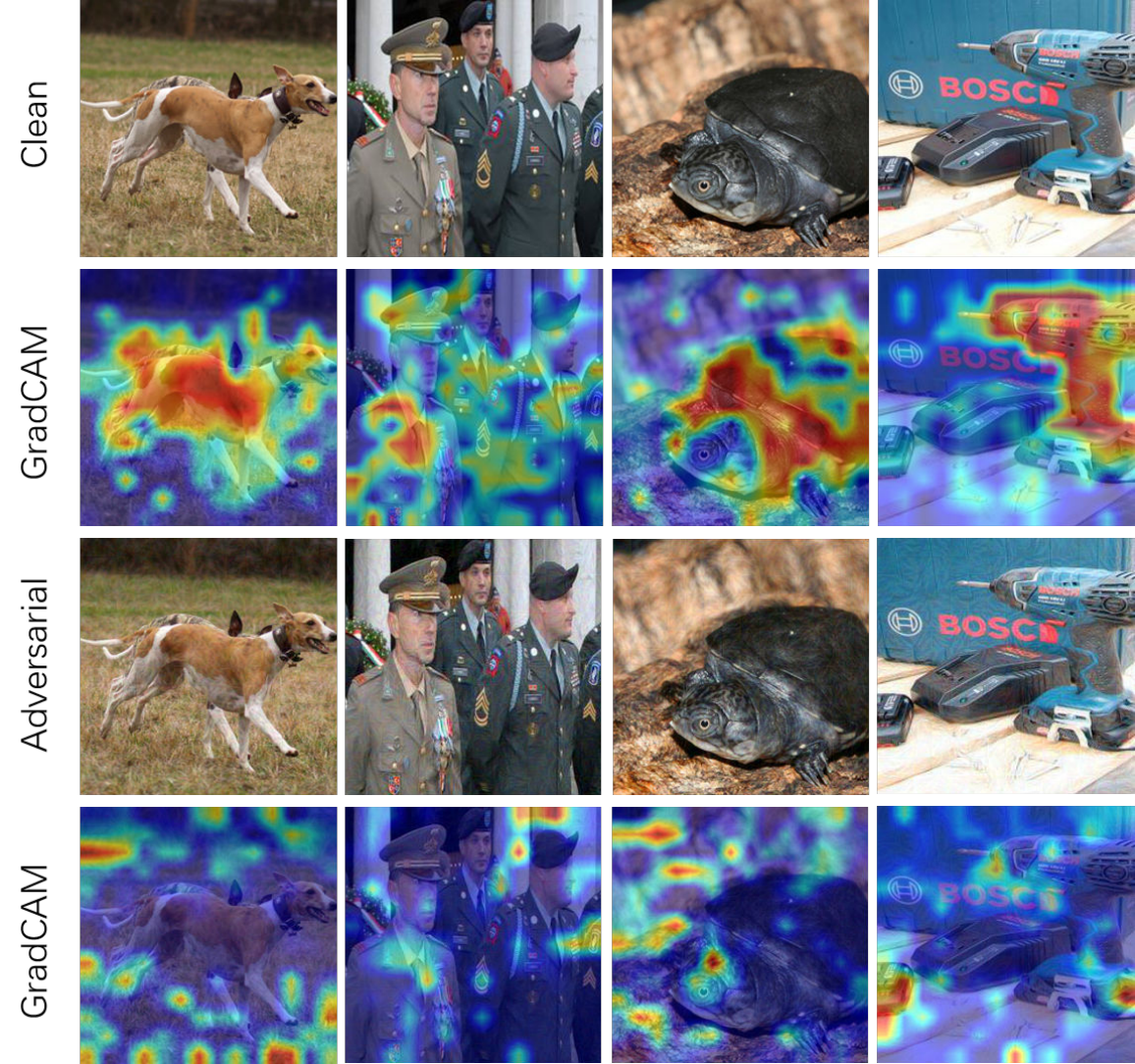}
\end{center}
   \caption{Grad-CAM for clean and adversarial samples.}
\label{fig:gradcam}
\end{figure}

We present visualizations of the Grad-CAM maps for 4 pairs of clean images and their corresponding adversarial examples in Figure~\ref{fig:gradcam}. Our analysis reveals that the Grad-CAM maps tend to behave in a more uniform and focused manner when presented with adversarial examples,with some instances showing a notable shift in focus.

\section{Conclusion}

In this paper, we present a label-free white-box attack approach for ViT-based models that exhibits strong transferability to various black-box models. Our inspiration comes from the feature collapse phenomenon in ViTs, 
causing the features in middle-to-end layers to become increasingly similar and eventually collapse. We propose the feature diversity attacker to naturally accelerate this process and achieve remarkable performance and transferability. 


There are several limitations of our approach, for example, our model relies on the model parameters and structure for generating attacking samples. We plan to tackle these problems in the future search.


\section*{Ethical Statement}
We state that the knowledge and discussion of such attacks will be used for defensive purposes—to improve the robustness and security of AI systems—not to perpetrate harm or malicious activity.

\section*{Acknowledgements}
This work is supported by the National Natural Science Foundation of China (NSFC No. 62272184). The computation is completed in the HPC Platform of Huazhong University of Science and Technology.

{
    \bibliography{aaai24}

\begin{thebibliography}{39}
\providecommand{\natexlab}[1]{#1}

\bibitem[{Barannyk et~al.(2015)Barannyk, Aboutaleb, Elshabini, and Barlow}]{barannyk2015spectrally}
Barannyk, L.~L.; Aboutaleb, H.~A.; Elshabini, A.; and Barlow, F.~D. 2015.
\newblock Spectrally accurate causality enforcement using SVD-based Fourier continuations for high-speed digital interconnects.
\newblock \emph{IEEE Transactions on Components, Packaging and Manufacturing Technology}, 5(7): 991--1005.

\bibitem[{Beal et~al.(2020)Beal, Kim, Tzeng, Park, Zhai, and Kislyuk}]{beal2020toward}
Beal, J.; Kim, E.; Tzeng, E.; Park, D.~H.; Zhai, A.; and Kislyuk, D. 2020.
\newblock Toward transformer-based object detection.
\newblock \emph{arXiv preprint arXiv:2012.09958}.

\bibitem[{Brown et~al.(2017)Brown, Man{\'e}, Roy, Abadi, and Gilmer}]{brown2017adversarial}
Brown, T.~B.; Man{\'e}, D.; Roy, A.; Abadi, M.; and Gilmer, J. 2017.
\newblock Adversarial patch.
\newblock \emph{arXiv preprint arXiv:1712.09665}.

\bibitem[{Carion et~al.(2020)Carion, Massa, Synnaeve, Usunier, Kirillov, and Zagoruyko}]{carion2020end}
Carion, N.; Massa, F.; Synnaeve, G.; Usunier, N.; Kirillov, A.; and Zagoruyko, S. 2020.
\newblock End-to-end object detection with transformers.
\newblock In \emph{Computer Vision--ECCV 2020: 16th European Conference, Glasgow, UK, August 23--28, 2020, Proceedings, Part I 16}, 213--229. Springer.

\bibitem[{Chen et~al.(2022)Chen, Zhang, Cheng, Awadallah, and Wang}]{chen2022principle}
Chen, T.; Zhang, Z.; Cheng, Y.; Awadallah, A.; and Wang, Z. 2022.
\newblock The principle of diversity: Training stronger vision transformers calls for reducing all levels of redundancy.
\newblock In \emph{Proceedings of the IEEE/CVF Conference on Computer Vision and Pattern Recognition}, 12020--12030.

\bibitem[{Dong, Cordonnier, and Loukas(2021)}]{dong2021attention}
Dong, Y.; Cordonnier, J.-B.; and Loukas, A. 2021.
\newblock Attention is not all you need: Pure attention loses rank doubly exponentially with depth.
\newblock In \emph{International Conference on Machine Learning}, 2793--2803. PMLR.

\bibitem[{Dong et~al.(2018)Dong, Liao, Pang, Su, Zhu, Hu, and Li}]{dong2018boosting}
Dong, Y.; Liao, F.; Pang, T.; Su, H.; Zhu, J.; Hu, X.; and Li, J. 2018.
\newblock Boosting adversarial attacks with momentum.
\newblock In \emph{Proceedings of the IEEE conference on computer vision and pattern recognition}, 9185--9193.

\bibitem[{Fu et~al.(2022)Fu, Zhang, Wu, Wan, and Lin}]{fu2022patch}
Fu, Y.; Zhang, S.; Wu, S.; Wan, C.; and Lin, Y. 2022.
\newblock Patch-fool: Are vision transformers always robust against adversarial perturbations?
\newblock \emph{arXiv preprint arXiv:2203.08392}.

\bibitem[{Gao et~al.(2020)Gao, Zhang, Song, Liu, and Shen}]{gao2020patch}
Gao, L.; Zhang, Q.; Song, J.; Liu, X.; and Shen, H.~T. 2020.
\newblock Patch-wise attack for fooling deep neural network.
\newblock In \emph{Computer Vision--ECCV 2020: 16th European Conference, Glasgow, UK, August 23--28, 2020, Proceedings, Part XXVIII 16}, 307--322. Springer.

\bibitem[{Gong et~al.(2021)Gong, Wang, Li, Chandra, and Liu}]{gong2021vision}
Gong, C.; Wang, D.; Li, M.; Chandra, V.; and Liu, Q. 2021.
\newblock Vision transformers with patch diversification.
\newblock \emph{arXiv preprint arXiv:2104.12753}.

\bibitem[{Goodfellow, Shlens, and Szegedy(2014)}]{goodfellow2014explaining}
Goodfellow, I.~J.; Shlens, J.; and Szegedy, C. 2014.
\newblock Explaining and harnessing adversarial examples.
\newblock \emph{arXiv preprint arXiv:1412.6572}.

\bibitem[{Hatamizadeh et~al.(2022)Hatamizadeh, Yin, Roth, Li, Kautz, Xu, and Molchanov}]{hatamizadeh2022gradvit}
Hatamizadeh, A.; Yin, H.; Roth, H.~R.; Li, W.; Kautz, J.; Xu, D.; and Molchanov, P. 2022.
\newblock Gradvit: Gradient inversion of vision transformers.
\newblock In \emph{Proceedings of the IEEE/CVF Conference on Computer Vision and Pattern Recognition}, 10021--10030.

\bibitem[{Huybrechs(2010)}]{huybrechs2010fourier}
Huybrechs, D. 2010.
\newblock On the Fourier extension of nonperiodic functions.
\newblock \emph{SIAM Journal on Numerical Analysis}, 47(6): 4326--4355.

\bibitem[{Inkawhich et~al.(2019)Inkawhich, Wen, Li, and Chen}]{inkawhich2019feature}
Inkawhich, N.; Wen, W.; Li, H.~H.; and Chen, Y. 2019.
\newblock Feature space perturbations yield more transferable adversarial examples.
\newblock In \emph{Proceedings of the IEEE/CVF Conference on Computer Vision and Pattern Recognition}, 7066--7074.

\bibitem[{Kim(2020)}]{kim2020torchattacks}
Kim, H. 2020.
\newblock Torchattacks: A pytorch repository for adversarial attacks.
\newblock \emph{arXiv preprint arXiv:2010.01950}.

\bibitem[{Kurakin, Goodfellow, and Bengio(2016)}]{kurakin2016adversarial}
Kurakin, A.; Goodfellow, I.; and Bengio, S. 2016.
\newblock Adversarial machine learning at scale.
\newblock \emph{arXiv preprint arXiv:1611.01236}.

\bibitem[{Lin et~al.(2019)Lin, Song, He, Wang, and Hopcroft}]{lin2019nesterov}
Lin, J.; Song, C.; He, K.; Wang, L.; and Hopcroft, J.~E. 2019.
\newblock Nesterov accelerated gradient and scale invariance for adversarial attacks.
\newblock \emph{arXiv preprint arXiv:1908.06281}.

\bibitem[{Lovisotto et~al.(2022)Lovisotto, Finnie, Munoz, Mummadi, and Metzen}]{lovisotto2022give}
Lovisotto, G.; Finnie, N.; Munoz, M.; Mummadi, C.~K.; and Metzen, J.~H. 2022.
\newblock Give me your attention: Dot-product attention considered harmful for adversarial patch robustness.
\newblock In \emph{Proceedings of the IEEE/CVF Conference on Computer Vision and Pattern Recognition}, 15234--15243.

\bibitem[{Mahmood, Mahmood, and Van~Dijk(2021)}]{mahmood2021robustness}
Mahmood, K.; Mahmood, R.; and Van~Dijk, M. 2021.
\newblock On the robustness of vision transformers to adversarial examples.
\newblock In \emph{Proceedings of the IEEE/CVF International Conference on Computer Vision}, 7838--7847.

\bibitem[{Naseer et~al.(2021)Naseer, Ranasinghe, Khan, Khan, and Porikli}]{naseer2021improving}
Naseer, M.; Ranasinghe, K.; Khan, S.; Khan, F.~S.; and Porikli, F. 2021.
\newblock On improving adversarial transferability of vision transformers.
\newblock \emph{arXiv preprint arXiv:2106.04169}.

\bibitem[{Raghu et~al.(2021)Raghu, Unterthiner, Kornblith, Zhang, and Dosovitskiy}]{raghu2021vision}
Raghu, M.; Unterthiner, T.; Kornblith, S.; Zhang, C.; and Dosovitskiy, A. 2021.
\newblock Do vision transformers see like convolutional neural networks?
\newblock \emph{Advances in Neural Information Processing Systems}, 34: 12116--12128.

\bibitem[{Russakovsky et~al.(2015)Russakovsky, Deng, Su, Krause, Satheesh, Ma, Huang, Karpathy, Khosla, Bernstein et~al.}]{russakovsky2015imagenet}
Russakovsky, O.; Deng, J.; Su, H.; Krause, J.; Satheesh, S.; Ma, S.; Huang, Z.; Karpathy, A.; Khosla, A.; Bernstein, M.; et~al. 2015.
\newblock Imagenet large scale visual recognition challenge.
\newblock \emph{International journal of computer vision}, 115: 211--252.

\bibitem[{Said et~al.(2023)Said, Atri, Albahar, Ben~Atitallah, and Alsariera}]{said2023scene}
Said, Y.; Atri, M.; Albahar, M.~A.; Ben~Atitallah, A.; and Alsariera, Y.~A. 2023.
\newblock Scene Recognition for Visually-Impaired People’s Navigation Assistance Based on Vision Transformer with Dual Multiscale Attention.
\newblock \emph{Mathematics}, 11(5): 1127.

\bibitem[{Shi and Han(2021)}]{shi2021decision}
Shi, Y.; and Han, Y. 2021.
\newblock Decision-based black-box attack against vision transformers via patch-wise adversarial removal.
\newblock \emph{arXiv preprint arXiv:2112.03492}.

\bibitem[{Strudel et~al.(2021)Strudel, Garcia, Laptev, and Schmid}]{strudel2021segmenter}
Strudel, R.; Garcia, R.; Laptev, I.; and Schmid, C. 2021.
\newblock Segmenter: Transformer for semantic segmentation.
\newblock In \emph{Proceedings of the IEEE/CVF international conference on computer vision}, 7262--7272.

\bibitem[{Tang et~al.(2021)Tang, Han, Xu, Xiao, Deng, Xu, and Wang}]{tang2021augmented}
Tang, Y.; Han, K.; Xu, C.; Xiao, A.; Deng, Y.; Xu, C.; and Wang, Y. 2021.
\newblock Augmented shortcuts for vision transformers.
\newblock \emph{Advances in Neural Information Processing Systems}, 34: 15316--15327.

\bibitem[{Wang et~al.(2022{\natexlab{a}})Wang, Zheng, Chen, and Wang}]{wang2022anti}
Wang, P.; Zheng, W.; Chen, T.; and Wang, Z. 2022{\natexlab{a}}.
\newblock Anti-oversmoothing in deep vision transformers via the fourier domain analysis: From theory to practice.
\newblock \emph{arXiv preprint arXiv:2203.05962}.

\bibitem[{Wang et~al.(2022{\natexlab{b}})Wang, Wang, Yin, Gong, Wang, Liu, and Liu}]{wang2022generating}
Wang, Y.; Wang, J.; Yin, Z.; Gong, R.; Wang, J.; Liu, A.; and Liu, X. 2022{\natexlab{b}}.
\newblock Generating transferable adversarial examples against vision transformers.
\newblock In \emph{Proceedings of the 30th ACM International Conference on Multimedia}, 5181--5190.

\bibitem[{Wei et~al.(2022)Wei, Chen, Goldblum, Wu, Goldstein, and Jiang}]{wei2022towards}
Wei, Z.; Chen, J.; Goldblum, M.; Wu, Z.; Goldstein, T.; and Jiang, Y.-G. 2022.
\newblock Towards transferable adversarial attacks on vision transformers.
\newblock In \emph{Proceedings of the AAAI Conference on Artificial Intelligence}, volume~36, 2668--2676.

\bibitem[{Wu et~al.(2020)Wu, Su, Chen, Zhao, King, Lyu, and Tai}]{wu2020boosting}
Wu, W.; Su, Y.; Chen, X.; Zhao, S.; King, I.; Lyu, M.~R.; and Tai, Y.-W. 2020.
\newblock Boosting the transferability of adversarial samples via attention.
\newblock In \emph{Proceedings of the IEEE/CVF Conference on Computer Vision and Pattern Recognition}, 1161--1170.

\bibitem[{Wu et~al.(2022)Wu, Liu, Zhan, and Cheng}]{wu2022p2t}
Wu, Y.-H.; Liu, Y.; Zhan, X.; and Cheng, M.-M. 2022.
\newblock P2T: Pyramid pooling transformer for scene understanding.
\newblock \emph{IEEE Transactions on Pattern Analysis and Machine Intelligence}.

\bibitem[{Xie et~al.(2019)Xie, Zhang, Zhou, Bai, Wang, Ren, and Yuille}]{xie2019improving}
Xie, C.; Zhang, Z.; Zhou, Y.; Bai, S.; Wang, J.; Ren, Z.; and Yuille, A.~L. 2019.
\newblock Improving transferability of adversarial examples with input diversity.
\newblock In \emph{Proceedings of the IEEE/CVF Conference on Computer Vision and Pattern Recognition}, 2730--2739.

\bibitem[{Xu et~al.(2022)Xu, Zhang, Zhang, and Tao}]{xu2022vitpose}
Xu, Y.; Zhang, J.; Zhang, Q.; and Tao, D. 2022.
\newblock Vitpose: Simple vision transformer baselines for human pose estimation.
\newblock \emph{arXiv preprint arXiv:2204.12484}.

\bibitem[{Ye, Cheng, and Huang(2022)}]{ye2022fg}
Ye, Z.; Cheng, X.; and Huang, X. 2022.
\newblock FG-UAP: Feature-Gathering Universal Adversarial Perturbation.
\newblock \emph{arXiv preprint arXiv:2209.13113}.

\bibitem[{Zhang et~al.(2023)Zhang, Huang, Wu, and Lyu}]{zhang2023transferable}
Zhang, J.; Huang, Y.; Wu, W.; and Lyu, M.~R. 2023.
\newblock Transferable Adversarial Attacks on Vision Transformers with Token Gradient Regularization.
\newblock In \emph{Proceedings of the IEEE/CVF Conference on Computer Vision and Pattern Recognition}, 16415--16424.

\bibitem[{Zheng et~al.(2021)Zheng, Zhu, Mendieta, Yang, Chen, and Ding}]{zheng20213d}
Zheng, C.; Zhu, S.; Mendieta, M.; Yang, T.; Chen, C.; and Ding, Z. 2021.
\newblock 3d human pose estimation with spatial and temporal transformers.
\newblock In \emph{Proceedings of the IEEE/CVF International Conference on Computer Vision}, 11656--11665.

\bibitem[{Zhou et~al.(2021)Zhou, Kang, Jin, Yang, Lian, Jiang, Hou, and Feng}]{zhou2021deepvit}
Zhou, D.; Kang, B.; Jin, X.; Yang, L.; Lian, X.; Jiang, Z.; Hou, Q.; and Feng, J. 2021.
\newblock Deepvit: Towards deeper vision transformer.
\newblock \emph{arXiv preprint arXiv:2103.11886}.

\bibitem[{Zhu et~al.(2021)Zhu, Zhu, Zhang, Wu, Fu, and Li}]{zhu2021unified}
Zhu, F.; Zhu, Y.; Zhang, L.; Wu, C.; Fu, Y.; and Li, M. 2021.
\newblock A unified efficient pyramid transformer for semantic segmentation.
\newblock In \emph{Proceedings of the IEEE/CVF International Conference on Computer Vision}, 2667--2677.

\bibitem[{Zhu et~al.(2022)Zhu, Chen, Li, Chen, He, Tian, Zheng, Chen, and Huang}]{zhu2022toward}
Zhu, Y.; Chen, Y.; Li, X.; Chen, K.; He, Y.; Tian, X.; Zheng, B.; Chen, Y.; and Huang, Q. 2022.
\newblock Toward Understanding and Boosting Adversarial Transferability From a Distribution Perspective.
\newblock \emph{IEEE Transactions on Image Processing}, 31: 6487--6501.

\end{thebibliography}
}

\end{document}